\documentclass[aps,prb,floatfix,twocolumn,tightenlines,amsmath,amssymb]{revtex4}

\usepackage{graphicx}

\usepackage{amsmath}

\usepackage{amssymb}

\newcommand*{\rom}[1]{\expandafter\@slowromancap\romannumeral #1@}

\usepackage{epstopdf}

\usepackage{color}

\newcommand{\ket}[1] {|#1 \rangle}

\newcommand{\ketbra}[2] {{|#1 \rangle\!\langle #2|}}

\begin{document}

\title{The impact of nuclear spin dynamics on electron transport through donors}
\author{S. K. Gorman, M. A. Broome, W. J. Baker, M. Y. Simmons}
\affiliation{Centre of Excellence for Quantum Computation and Communication Technology, School of Physics, University of New South Wales, Sydney, New South Wales 2052, Australia}
\date{\today}

\begin{abstract}
We present an analysis of electron transport through two weakly coupled precision placed phosphorus donors in silicon. In particular, we examine the (1,1)$\leftrightarrow$(0,2) charge transition where we predict a new type of current blockade driven entirely by the nuclear spin dynamics. Using this nuclear spin blockade mechanism we devise a protocol to readout the state of single nuclear spins using electron transport measurements only. We extend our model to include realistic effects such as Stark shifted hyperfine interactions and multi-donor clusters. In the case of multi-donor clusters we show how nuclear spin blockade can be alleviated allowing for low magnetic field electron spin measurements.

\end{abstract}

\maketitle

\section{Introduction}

An understanding of electron transport through multiple quantum dots has enabled the progression of semiconductor quantum information protocols from single shot spin readout to two-qubit logic gates~\cite{johnson2005,koppens2005,koppens2006,nowack2011}. Not only do transport measurements provide us with important details on spin relaxation times and tunnel rates, but they play a vital role in aiding our understanding of the complex spin dynamics that occur in these systems, such as coherent manipulation of the electron spins~\cite{koppens2005,petta2005} and dynamical nuclear polarisation of the Overhauser field~\cite{schuetz2014}. 

With recent advances in the fabrication of precision placed donors in silicon~\cite{prati2012,gonzalez-zalba2014,fuechsle2012,weber2012a,buch2013} research in this field is now focused on spin transport through multi-donor chains. A deeper understanding of the interplay between electron- and nuclear-spins in the dynamics of such systems is a prerequisite for the progression of this field. In particular for the implementation of spin transport via donor chains~\cite{hollenberg2006}. So far, protocols based on spin-chains~\cite{bose2003,bose2007,kay2010}, coherent tunneling adiabatic passage (CTAP)~\cite{hollenberg2006,rahman2009,rahman2010}, spin shuttling~\cite{cirac2000,skinner2003} and SWAP-gate operations~\cite{loss1998} have been proposed, all of which require control of electron spin transport across donors. In order to further investigate these transport protocols it is crucial to understand the spin dynamics at the two donor level. However, despite the plethora of theoretical knowledge on gate-defined semiconductor double quantum dots~\cite{vanderwiel2003,taylor2007,hanson2007}; double donor transport has not received as much attention.

In this paper we show how the interplay between the electron and nuclear spins in donor-based systems affects not only the charge transport, but also the spin transport. To understand the impact of these nuclear spins we use a master equation approach to conduct a comprehensive numerical analysis of electron transport through a double donor system. We investigate electron spin resonance (ESR) combined with Pauli spin blockade (PSB) at low and high magnetic fields to manipulate and readout the electron spin states. Our most striking finding demonstrates that the presence of the quantised nuclear spin of the donors leads to a novel effect called nuclear spin blockade. Using this mechanism we propose a new spin readout protocol for the nuclear spins based on a measurement of the transport current. Finally, we analyse more realistic scenarios of inhomogeneous hyperfine interactions across the donors, as well as the case of multi-donor dots, which can be shown to be immune to nuclear spin blockade. Throughout the paper we neglect the dynamical behavior of the surrounding \textsuperscript{29}Si nuclear spins present in natural silicon. This interaction is much smaller than the donor hyperfine interaction~\cite{schliemann2003} and it has also been shown that Si:P devices can be fabricated in isotopically pure \textsuperscript{28}Si where the absence of the \textsuperscript{29}Si extends the electron coherence times~\cite{muhonen2014}.

\section{Transport at the (1,1) to (0,2) charge transition}
\begin{figure}
\begin{center}
\includegraphics[width=1\columnwidth]{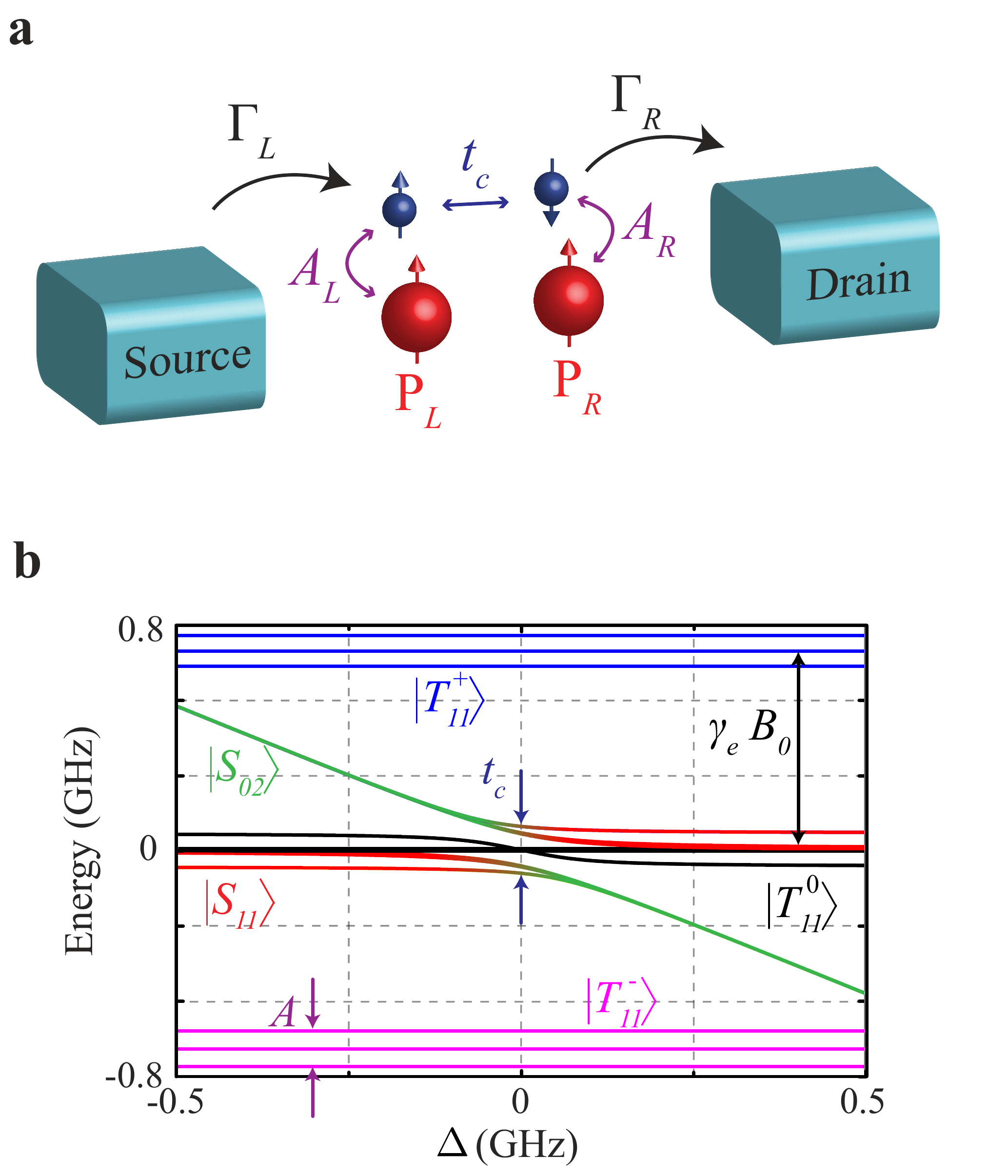}
\end{center}
\caption{{\bf Electron transport through two weakly coupled phosphorus donors in silicon.} {\bf (a)} A schematic representation of transport through a double donor system during PSB. Electrons can tunnel from the source to P$_L$ and from P$_R$ to the drain at rates $\Gamma_L$ and $\Gamma_R$, respectively. The two donor electrons are coherently tunnel coupled at the rate $t_c$ and have a contact hyperfine interaction with the nuclear spins, $A_L$ and $A_R$ with their respective nuclei. {\bf (b)} Eigen energies of $H$ around $\Delta{=}0$, between the (1,1) and (0,2) charge configurations at an external magnetic field of $B_0{=}25$~mT and with hyperfine interaction strength set to $A_L{=}A_R{=}A{=}117.53$~MHz. The electron triplet states, $\ket{T_{11}^+}$ (blue), $\ket{T_{11}^0}$ (black), and $\ket{T_{11}^-}$ (pink) are split by the Zeeman energy, $\gamma_e B_0$. The $\ket{S_{02}}$ (green) state is detuned from the $\ket{S_{11}}$ (red) with an anti-crossing at $\Delta{=}0$ due to a tunnel coupling set to $t_c=A$. The (0,1) charge states are omitted for clarity.}
\label{fig:intro}
\end{figure}

We consider two weakly coupled phosphorus donors in a silicon lattice (approximately 15--20 nm apart~\cite{koiller2002,wellard2003}), P$_L$ and P$_R$ the left and right donor, respectively. Electrons are able to tunnel from in plane source to drain leads via both donors as shown schematically in Fig.~\ref{fig:intro}a. Describing the system is the Hamiltonian $H$, 
\begin{equation}
H = H_{ze} + H_{zn} + H_{t_c} + H_{\Delta} + H_{hf},
\end{equation}
where $H_{ze}$ and $H_{zn}$ are the electron and nuclear Zeeman terms, $H_{t_c}$ is the tunnel coupling between the donor electrons, $H_{\Delta}$ is the energy detuning of the $\ket{S_{02}}$ state (singlet state with two electrons on a single donor nuclei) and $H_{hf}$ is the hyperfine interaction, for further details see Methods. Throughout the paper we refer to the Hamiltonian in the singlet-triplet basis of the electrons with energies shown in Fig.~\ref{fig:intro}b. By making a transformation from Hilbert space to Liouville space we incorporate incoherent processes that occur during spin transport~\cite{ernst1987}. In doing so, tunnelling that occurs from the source at the rate $\Gamma_L$, to the donors and through to the drain at the rate $\Gamma_R$, are integrated with the coherent evolution of the system. Using this approach we can determine the spin and charge dynamics of the donor system during electron transport.

As a first demonstration of the effect of quantised nuclear spin states we study the electron transport from drain to source (reverse of Fig.~\ref{fig:intro}a). In this scenario the charge cycle is (0,1){$\rightarrow$}(0,2){$\rightarrow$}(1,1){$\rightarrow$}(0,1), where ($n_L$, $n_R$) corresponds to the electron numbers on the left and right donor nuclei. In the case of quantum dots it has been shown that the current, $I_{QD}$, as a function of the detuning, $\Delta$, is given by a known expression involving the coherent and incoherent tunnel rates~\cite{nazarov1993},
\begin{equation}
I_{QD} = \frac{|e| \Gamma_L \left(\frac{t_c}{2}\right)^2}{\left(\frac{\Gamma_R}{2}\right)^2 + \left(\frac{t_c}{2}\right)^2(2 + \frac{\Gamma_L}{\Gamma_R}) + \Delta^2},
\end{equation}
where $e$ is the electron charge and $t_c$ is the tunnel coupling between the two dots. However, without an external magnetic ($B_0$) dependence, this equation will not account for any spin dynamics. In the case of donors, the measurable current is given by $I{=}\left|e\right|\Gamma_L P$(1,1), where $P$(1,1) is the probability of being in any (1,1) charge configuration, including the electron triplet states $\{\ket{T_{11}^+}, \ket{T_{11}^-} \}$.

\begin{figure}[b!]
\begin{center}
\includegraphics[width=1\columnwidth]{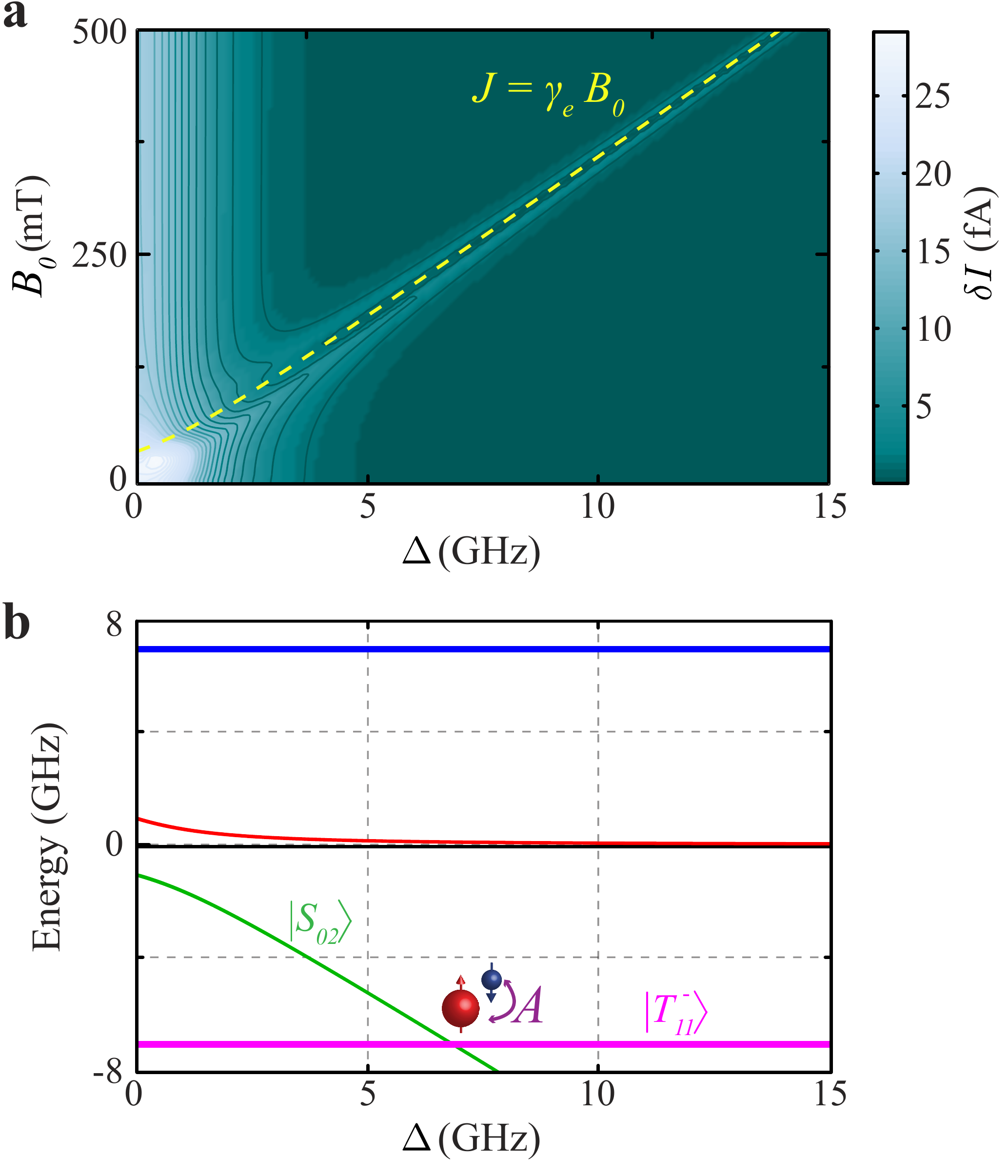}
\end{center}
\caption{{\bf Electron transport through double donors from drain to source.} {\bf (a)} The difference in current, $\delta I$, through a double donor as a function of detuning $\Delta$, and external magnetic field, $B_0$. The yellow line maps $J{=}\gamma_e B_0$ which follows a peak in $\delta I$ due to the mixing between $\ket{S_{02}}{\leftrightarrow}\ket{T_{11}^-}$ (contour lines guide the eye). {\bf (b)} Eigen energies of $H$ at $B_0{=}250$~mT as a function of the detuning, showing the $\ket{S_{02}}{\leftrightarrow}\ket{T_{11}^-}$ anti-crossing. Here, $H_{hf}$ allows for electron-nuclear spin flip-flops, increasing $P$(1,1). The (0,1) charge states are omitted for clarity. In this simulation $t_c{=}2$ GHz,  $A_L{=}A_R{=}A{=}117.53$ MHz, $\Gamma_L{=}\Gamma_R{=}100$ MHz.}
\label{fig:nazz}
\end{figure}

To investigate the effect the inclusion of these states has on electron transport we plot the difference in current, $\delta I{=}I{-}I_{QD}$, for $\Delta{\ge}0$ in Fig.~\ref{fig:nazz}a. There is a peak in the current that can be seen to follow the point at which the electron exchange interaction, $J{=}\Delta/2{+}\sqrt{(t_c/2)^2{+}(\Delta/2)^2}$, ($t_c{=}2$ GHz) is equal to the electron Zeeman energy, $\gamma_e B_0$. This is analogous to the canonical spin-funnel experiments~\cite{petta2005}. At this value of detuning the hyperfine interaction mixes electron singlet-triplet states $\ket{S_{02}}{\leftrightarrow}\ket{T_{11}^-}$ via a nuclear spin flip, see Fig.~\ref{fig:nazz}b. This increase in the (1,1) state population allows for the electron on P$_L$ to tunnel off to the source giving rise to a larger current. In gate-defined quantum dots this current peak is not observed since the large nuclear spin bath has a distribution of hyperfine strengths. Therefore, the position of this peak shown in Fig.~\ref{fig:nazz}a, will shift depending on the exact nuclear spin configuration and will be averaged out due to the fluctuating Overhauser field~\cite{jouravlev2006}.

For $\Delta{<<}0$ transport across the donors becomes energetically unfavourable as the electrons will remain in the (1,1) region without first loading into the (0,2) charge state, see Fig.\ref{fig:intro}b. We therefore do not expect to see mixing between $\ket{S_{02}}{\leftrightarrow}\ket{T_{11}^+}$ and a current peak will not be observed. 

\section{Nuclear spin blockade}
Next, we consider the transport cycle from source to drain: (0,1){$\rightarrow$}(1,1){$\rightarrow$}(0,2){$\rightarrow$}(0,1) where we can expect PSB due to the large energy splitting of the (0,2) electron singlet-triplet states~\cite{ono2002}. As a consequence, if the electrons are in any of the (1,1) triplet states an electron on P$_L$ cannot tunnel to P$_R$. Therefore, tunneling can only occur via the singlet states and the current is given solely by the $\ket{S_{02}}$ probability: $I{=}\left|e\right|\Gamma_R P(\ket{S_{02}})$.

It has been shown; however, that by applying a continuous-wave ESR magnetic field PSB can be lifted by driving transitions within the (1,1) electron triplet manifold, $\{\ket{T_{11}^+},\ket{T_{11}^-}\}{\leftrightarrow}\ket{T_{11}^0}$~\cite{koppens2006,koppens2007}. This applies when the tunnel coupling, $t_c$, is on the order $A$, hence, for the following analysis we set $t_c{=}A$. Importantly, any difference in the Larmor frequencies of the two electrons, $\delta\omega_e$, will result in $\ket{T^0_{11}}{\leftrightarrow}\ket{S_{11}}$ transitions~\cite{koppens2006}. The $\ket{S_{11}}$ state can in turn tunnel to $\ket{S_{02}}$ via the tunnel coupling $t_c$, allowing current to flow. In the presence of a homogeneous $B_0$ field $\delta\omega_e$ can only come from a difference in the nuclear spin orientation between the two donors.

We examine the spin transport by initialising in the (0,1) charge configuration and preparing a fully mixed state in both the nuclear and electron spins states such that at $t=0$, $\rho_0{=}({\ketbra{{\uparrow_{01}}}{{\uparrow_{01}}}} + {\ketbra{{\downarrow_{01}}}{{\downarrow_{01}}})} \otimes {\{N\}}$ (normalisation omitted). We first let the system reach PSB after which a continuous wave ESR field is applied with strength $B_1$ and on resonance at the frequency,
\begin{equation}
\Omega = (\gamma_e - \gamma_n) B_0 + \frac{t_c}{4} + \frac{A_L - A_R}{2} + \frac{\sqrt{\frac{(A_L + A_R)^2}{4} + \frac{t_c^2}{4}}}{2},
\end{equation}

\begin{figure}
\begin{center}
\includegraphics[width=1\columnwidth]{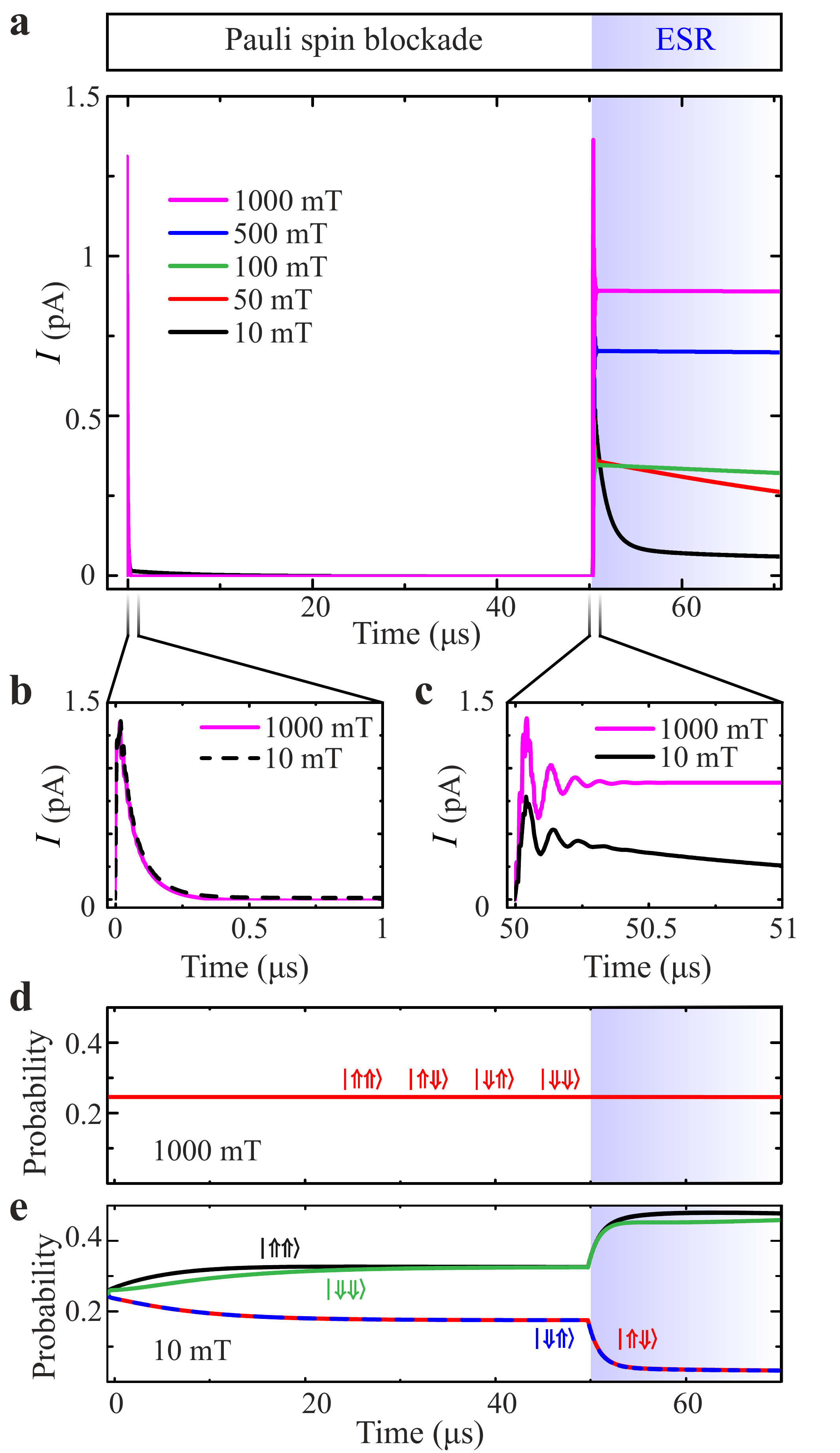}
\end{center}
\caption{{\bf Current during Pauli spin blockade and nuclear spin blockade.} {\bf (a)} The simulated current through two weakly coupled donors in transport for different external magnetic fields, $B_0$. At $t{=}0$ the density operator is given by $\rho_0{=}({\ketbra{{\uparrow_{01}}}{{\uparrow_{01}}}} + {\ketbra{{\downarrow_{01}}}{{\downarrow_{01}}})} \otimes {\{N\}}$ (normalisation omitted). {\bf (b)} PSB is reached in $\sim 0.1$ $\mu$s and the current drops to zero. {\bf (c)} An ESR driving field is turned on after 50 $\mu$s affecting the spin of both donor electrons. For all values of $B_0$ there is an initial spike in the current corresponding to the lifting of PSB followed by coherent oscillations. {\bf (d)} The nuclear spin state projections at $B_0{=}1000$~mT are unchanged during electron transport or after application of ESR due to the reduced electron-nuclear flip-flop events. {\bf (e)} At $B_0{=}10$~mT the nuclear $\ket{{\Uparrow \Downarrow}}$ and $\ket{{\Downarrow \Uparrow}}$ approach zero probability (${<}$0.05 after 50 $\mu$s), this is nuclear spin blockade. Simulations were carried out with $t_c{=}A_L{=}A_R{=}A{=}117.53$ MHz, $\Delta{=}0$, $\Gamma_L{=}\Gamma_R{=}100$ MHz.}
\label{fig:B0_sweep}
\end{figure}

\begin{figure*}[t!]
\begin{center}
\includegraphics[width=1\textwidth]{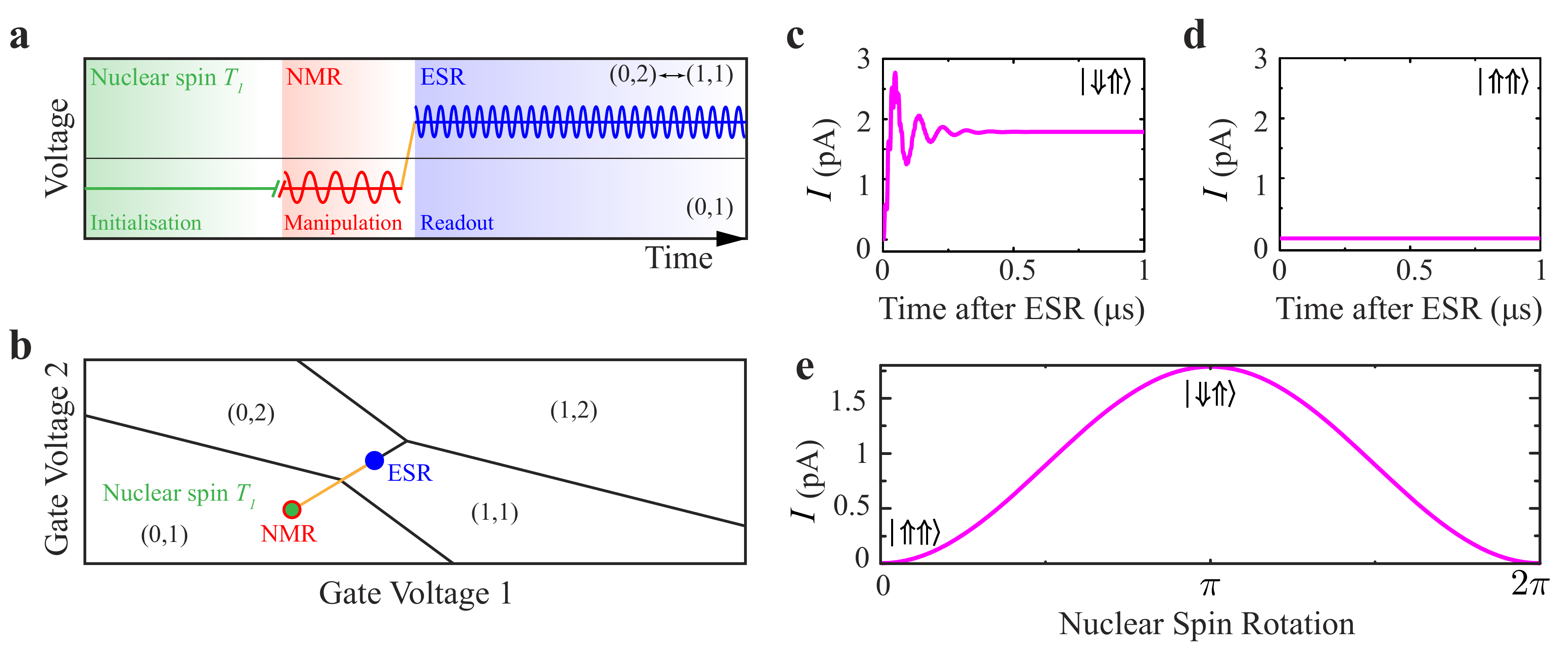}
\end{center}
\caption{{\bf Nuclear spin readout using two weakly coupled donors in transport.} {\bf (a)} Schematic representation of the proposed scheme for single nuclear spin readout of the P$_L$ nuclear spin showing gate pulsing, NMR and ESR driving. The nuclear spins are initialised in $\ket{{\Uparrow \Uparrow}}$ by waiting $T_1$. NMR can then be performed on P$_L$ nuclei by driving at a frequency $\gamma_n B_0$, which does not effect the P$_R$ nuclear spin due to the presence of the electron in the (0,1) charge state. {\bf (b)} The charge stability diagram for the double donor system with colors corresponding to movements in gate space for {\bf (a)}. Readout is performed by pulsing to the (1,1)$\leftrightarrow$(0,2) transition and applying ESR. {\bf (c)} If the nuclear spin state after NMR is $\ket{{\Downarrow \Uparrow}}$ then nuclear spin blockade does not occur. {\bf (d)} However, if the spin state is $\ket{{\Uparrow \Uparrow}}$ then no current will flow due to nuclear spin blockade. {\bf (e)} The current through the donors is linearly proportional to the spin down probability on P$_L$ allowing different NMR rotations to be mapped using the magnitude of the current during readout. Here, $B_0{=}1000$ mT, $t_c{=}A_L{=}A_R{=}A{=}117.53$ MHz, $\Delta{=}0$, $\Gamma_L{=}\Gamma_R{=}100$ MHz.}
\label{fig:readout}
\end{figure*}

\noindent where $\gamma_e$ and $\gamma_n$ are the electron and nuclear gyromagnetic ratios, and $A_L$ and $A_R$ are the nuclear hyperfine interactions on P$_L$ and P$_R$ respectively. 

At high magnetic fields, $B_0{\gtrsim}100$ mT, there is an initial peak in the current due to the tunneling of the electrons into the (1,1) charge region, see Fig.~\ref{fig:B0_sweep}a. In a time $\sim 0.1$ $\mu$s PSB is reached and the current is blocked (Fig.~\ref{fig:B0_sweep}b). After application of the ESR excitation the current initially spikes and coherent oscillations can be seen corresponding to electron spin rotations introducing singlet content at a frequency determined by the ESR field strength ($\gamma_e B_1 \sim 140$ MHz, Fig.~\ref{fig:B0_sweep}c). The current quickly finds a steady state centred around the oscillations from the ESR driving. During this time, the nuclear spins are unaffected by the electron transport. Figure~\ref{fig:B0_sweep}d shows the four nuclear spin state projections, $\{\ket{{\Uparrow \Uparrow}}, \ket{{\Uparrow \Downarrow}}, \ket{{\Downarrow \Uparrow}}, \ket{{\Downarrow \Downarrow}}\} = 0.25$.

At low magnetic fields, $B_0{<}100$ mT, the nuclear spin dynamics play a larger role in determining the current through the donors. Here the hyperfine interaction can cause electron-nuclear spin flip-flops ($\ket{{\uparrow \Downarrow}}{\leftrightarrow}\ket{{\downarrow \Uparrow}}$)~\cite{mccamey2009,mccamey2010}. This process is seen to increase the nuclear spin $\ket{{\Uparrow \Uparrow}}$ and $\ket{{\Downarrow \Downarrow}}$ populations to $\sim$0.3 whilst the system moves into PSB, see Fig.~\ref{fig:B0_sweep}e. The flip-flop process is further amplified when ESR is applied during which the nuclear $\ket{{\Uparrow \Downarrow}}$ and $\ket{{\Downarrow \Uparrow}}$ populations approach zero. Consequently, this reduces the mixing between $\ket{S_{11}}$ and $\ket{T_{11}^0}$ since $\delta\omega_e{\rightarrow}0$. Irrespective of the initial nuclear spin state, PSB cannot be lifted at low magnetic fields for double donor transport. We refer to this effect as ``nuclear spin blockade'' since the current through the donors is being blocked as a result of the nuclear spin states.

\section{Nuclear spin readout}
In the same way that PSB has been used to perform single electron spin readout in electron transport~\cite{koppens2006}, we investigated if nuclear spin blockade can be used to perform readout of a single nuclear spin. Even at high magnetic fields nuclear spin blockade prevents current flow if the nuclear spin states are aligned ($\ket{{\Uparrow \Uparrow}}$ or $\ket{{\Downarrow \Downarrow}}$) and allows current to flow if they are anti-aligned ($\ket{{\Uparrow \Downarrow}}$ or $\ket{{\Downarrow \Uparrow}}$). Importantly, at high fields the nuclear spin states remain unaffected by the electron transport; therefore, it is possible to use nuclear spin blockade as a readout mechanism for the nuclear spin states.

\begin{figure}
\begin{center}
\includegraphics[width=1\columnwidth,natwidth=610,natheight=642]{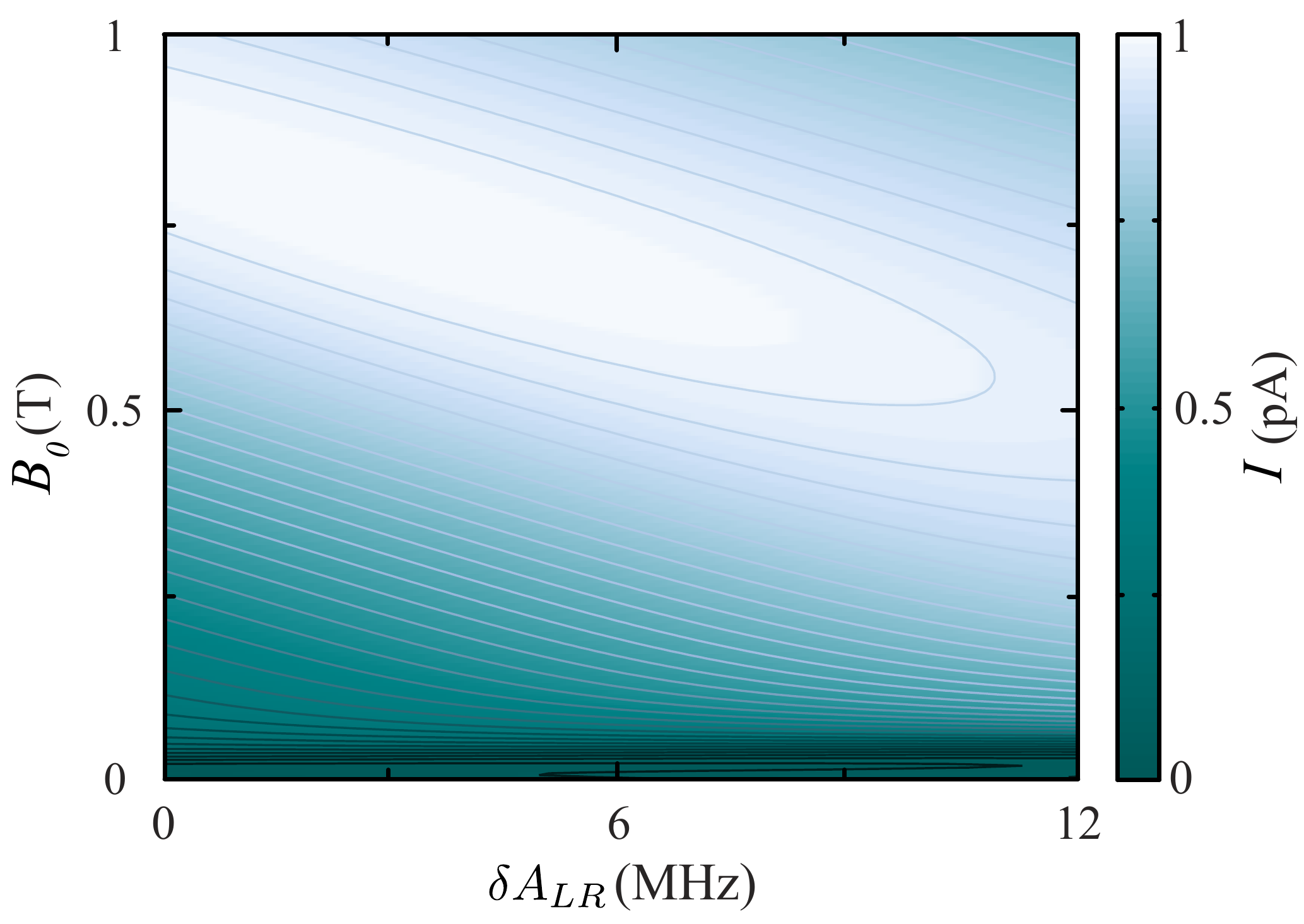}
\end{center}
\caption{{\bf Effect of inhomogeneous contact hyperfine interactions.} At high magnetic fields where the nuclear spin dynamics are unaffected by the electron transport the effect of larger $\delta A_{LR}{=}|A_L - A_R|$ reduces the current. In order to achieve the largest current there is an optimum $B_0$ for any given $\delta A_{LR}$ (contour lines guide the eye). Here, $\Delta{=}0$, $\Gamma_L{=}\Gamma_R{=}100$ MHz.}
\label{fig:dA}
\end{figure}

The spin measurement protocol consists of 3 stages: initialisation, manipulation and readout, see Fig.~\ref{fig:readout}a,b. We initialise in the (0,1) charge region by waiting for the $T_1$ of the nuclear spins, thereby preparing them in the ground state, $\ket{{\Uparrow \Uparrow}}$. The manipulation step involves applying NMR to the P$_L$ nuclear spin (which does not have an electron present) at a frequency $\gamma_n B_0$. Importantly, this will not effect the P$_R$ nuclear spin due to the presence of the donor electron. The readout consists of moving to the (1,1)$\rightarrow$(0,2) transition and applying ESR to test the presence of nuclear spin blockade. If the nuclear spin on P$_L$ is $\ket{{\Downarrow}}$ then current will flow through the donors (Fig.~\ref{fig:readout}c); however, if the nuclear spin is $\ket{{\Uparrow}}$ then the current will be blocked (Fig.~\ref{fig:readout}d). The current is also linearly proportional to the orientation of the nuclear spin; that is, if the total nuclear spin state is $(\ket{{\Downarrow \Uparrow}} + \ket{{\Uparrow \Uparrow}})/\sqrt{2}$ then the current will be half that of the $\ket{{\Downarrow \Uparrow}}$ state, see Fig.~\ref{fig:readout}e.

\section{Stark shift and donor clusters}
In this final section we extend the simulation work to examine non-idealised scenarios of Stark shifted contact hyperfine interactions and donor clusters (multi-donor quantum dots) in donor based transport, for example, in Ref.~\cite{weber2012a,weber2014,watson2014}.

\begin{figure}
\begin{center}
\includegraphics[width=1\columnwidth,natwidth=610,natheight=642]{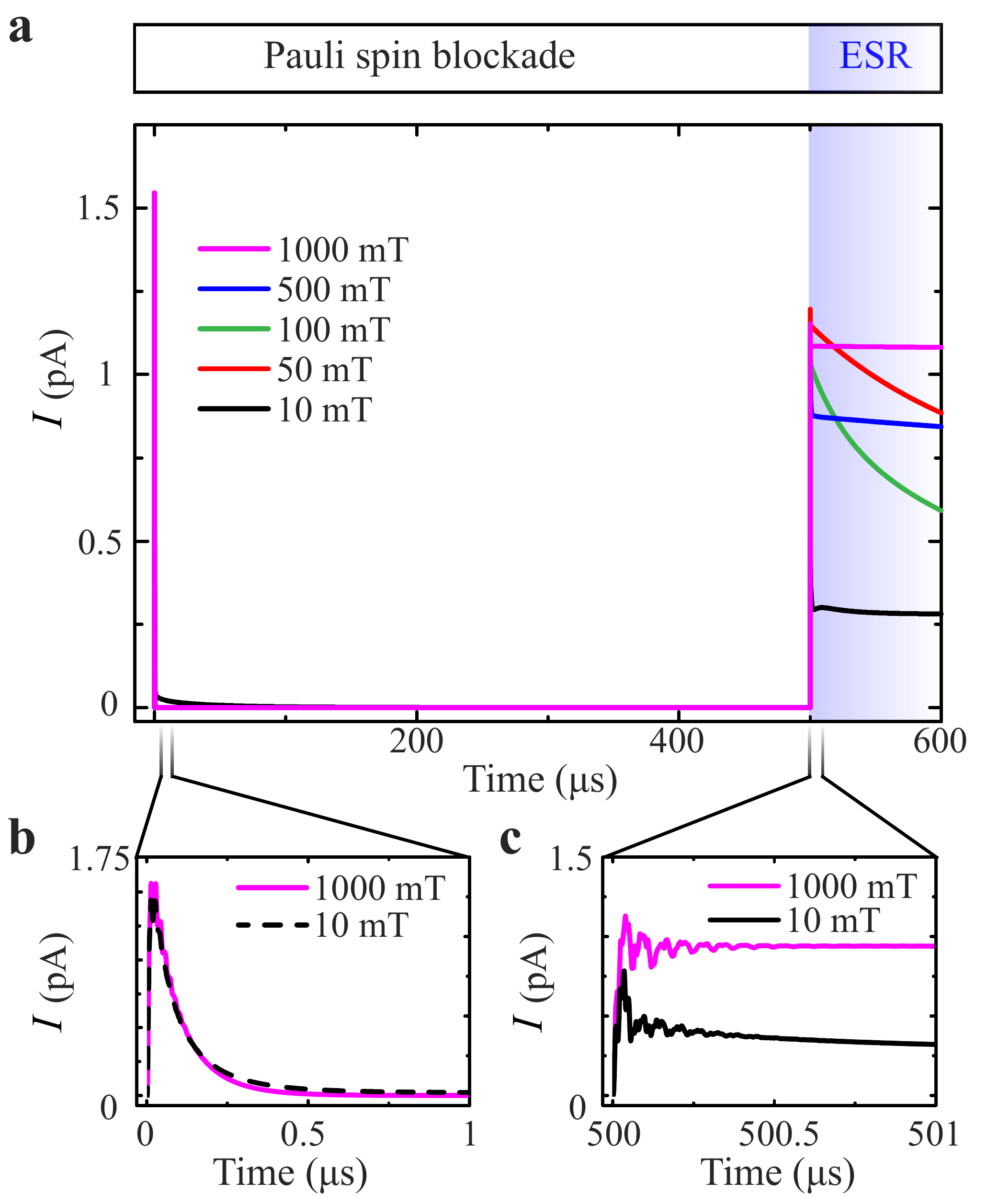}
\end{center}
\caption{{\bf Current through a single donor and a two-donor cluster.} {\bf (a)} Simulated electron transport for a single donor (P$_L$) and two donor cluster (2P$_R$) at different magnetic fields showing the absence of nuclear spin blockade. The initial density operator was $\rho_0{=}({\ketbra{{\uparrow_{01}}}{{\uparrow_{01}}}} + {\ketbra{{\downarrow_{01}}}{{\downarrow_{01}}})} \otimes {\{N_{1P/2P}\}}$ (normalisation omitted). {\bf (b)} PSB is reached on a similar time scale for two single donors in transport for all magnetic fields. {\bf (c)} After 500 $\mu$s an ESR driving field is applied to both electron spins. At low magnetic fields the current reaches a steady state but it does not decay to zero. The absence of decay in the current at low magnetic fields indicates that nuclear spin blockade is no longer present. At intermediate magnetic fields (50 mT and 100 mT) there is a decay in the current; however, the final steady state current is larger than that obtained for two single donors in transport. In this simulation, $t_c{=}A_L{=}A_R{=}117.53$ MHz, $\Delta{=}0$, $\Gamma_L{=}\Gamma_R{=}100$ MHz.}
\label{fig:1p_2p}
\end{figure}

The electron wavefunction around a donor nuclei can be distorted by electric fields which changes the contact hyperfine interaction, known as the Stark effect~\cite{buch2013,bradbury2006}. Due to the stray electric fields in the silicon crystal, e.g. from electronic gates or charges, it is most likely that two donors will naturally experience different hyperfine interactions. This inherent Stark shift is typically a few MHz in frequency~\cite{pla2012} and will create a difference in the hyperfine interaction between the two donors, $\delta A_{LR}{=}|A_L - A_R|$.

Figure~\ref{fig:dA} shows the steady state current after PSB and ESR (as in Fig.~\ref{fig:B0_sweep}a) for various magnetic fields and $\delta A_{LR}$ values. The current increases with the magnetic field up to a certain field and interestingly there is an optimum $B_0$ for any given $\delta A_{LR}$. At high magnetic fields the current decreases as $\delta A_{LR}$ becomes larger. This is due to $\delta A_{LR}$ decreasing $\delta\omega_e$ when the spins are anti-aligned, resulting in less $\ket{S_{11}}{\leftrightarrow}\ket{T_{11}^0}$ mixing and a lower current. At low magnetic fields the effect is reversed. Here, we are subject to nuclear spin blockade as previously discussed, however increasing $\delta A_{LR}$ will only increase $\delta\omega_e$ resulting in a higher current.

Using scanning tunneling microscopy (STM) hydrogen lithography it has been shown that donor clusters consisting of a small number of P atoms can be fabricated in Si that exhibit PSB~\cite{weber2014}. It is therefore possible to engineer a system consisting of a single donor and a two donor cluster in transport, essentially controlling the hyperfine interaction at the atomic scale. Now, in this system the electrons will always experience a $\delta\omega_e$ across the two dots, ensuring there is substantial mixing (${\gtrsim}A$) between the $\ket{T_{11}^0}$ and $\ket{S_{11}}$ states to lift PSB. This can be  seen if we consider the 8 possible nuclear configurations: $\{N_{1P/2P}\}{=}\{\ket{{\Uparrow}}\ket{{\Uparrow\Uparrow}}$, $\ket{{\Uparrow}}\ket{{\Uparrow\Downarrow}}$, $\ket{{\Uparrow}}\ket{{\Downarrow\Uparrow}}$, $\ket{{\Uparrow}}\ket{{\Downarrow\Downarrow}}$, $\ket{{\Downarrow}}\ket{{\Uparrow\Uparrow}}$, $\ket{{\Downarrow}}\ket{{\Uparrow\Downarrow}}$, $\ket{{\Downarrow}}\ket{{\Downarrow\Uparrow}}$, $\ket{{\Downarrow}}\ket{{\Downarrow\Downarrow}}\}$, where the first term is the spin state for P$_L$ and the second term with a two nuclear spin state is for 2P$_R$.

We simulate the current through a single donor (P$_L$) and two donor cluster (2P$_R$) in Fig.~\ref{fig:1p_2p}a at different magnetic fields with the initial state $\rho_0{=}({\ketbra{{\uparrow_{01}}}{{\uparrow_{01}}}} + {\ketbra{{\downarrow_{01}}}{{\downarrow_{01}}})} \otimes {\{N_{1P/2P}\}}$ at $t{=}0$. With this single and two donor cluster scenario, the nuclear spins take longer to reach a steady state configuration despite the electron spins reaching PSB in the same time scale as the single donor case, Fig.~\ref{fig:1p_2p}b. As a consequence we simulate the dynamics here for much longer. 

During ESR excitation the current is seen to oscillate with a beating due to the different hyperfine interactions experienced by each electron, see Fig.~\ref{fig:1p_2p}c. For all magnetic field strengths, the current is higher than it was in the single donor cases. Importantly, even at low magnetic fields, and in contrast to the single donor dynamics (Fig.~\ref{fig:B0_sweep}a), the current does not decay to zero. Instead, it finds a steady state solution centered around the coherent oscillations indicating that nuclear spin blockade is no longer present. 

\section{Summary}
We have developed a numerical model to investigate electron spin transport through donors in silicon. Our findings show that there are surprising effects that arise due to the quantised nature of the donor nuclei. We show how it is possible to map out the electron exchange interaction using simple transport measurements and predict a new current blockade mechanism as a consequence of nuclear spin dynamics. This is in contrast to other systems such as GaAs double quantum dots where large fluctuating nuclear spin baths averages out the quantum nature of the nuclear spins and these effects are not observed.

Importantly, we find that for low magnetic fields, $B_0{<}100$ mT, where long electron $T_1$ times~\cite{morello2010} and faster ESR rotations are possible, nuclear spin blockade prevents the transfer of the electron spins across the donors. Therefore, electron spin readout can only be performed at high magnetic fields; in which, the nuclear spin states are initialised in $\ket{{\Uparrow \Downarrow}}$ or $\ket{{\Downarrow \Uparrow}}$ otherwise $\delta\omega_e{=}0$ and there will be no $\ket{S_{11}}{\leftrightarrow}\ket{T_{11}^0}$ mixing. In addition, we demonstrate that there is an optimum $B_0$ field for a given Stark shift across the donors to achieve the maximum current after ESR driving.

Interestingly, we show how nuclear spin blockade can be utilised as a new readout mechanism for nuclear spins. This method, which only requires electron transport measurements and removes the need for traditional charge sensors will be a useful experimental tool for probing multi-donor interactions. The nuclear spin coherence times can be measured during electron shuttling which is critical for many proposed spin transport protocols~\cite{cirac2000,skinner2003}.

Finally, we demonstrate that for electron spin transport experiments where nuclear spin blockade is undesirable, multi-donor clusters can be used since they allow for lifting of PSB even at low magnetic fields. Advances in fabrication technologies, in particular STM lithography, offers the ability to tailor hyperfine interactions at the atomic scale for absolute control over the combined electron-nuclear spin system.

Our results provide important insights into the complex spin dynamics in double donor systems. Such an understanding is necessary for two-qubit interactions~\cite{kalra2014} and spin transport protocols in multi-donor chains~\cite{oh2013} for scalable solid-state quantum computing architectures~\cite{hollenberg2006}.

$ $\\
\noindent\textbf{Acknowledgments}
We thank T. L. Keevers and J. G. Keizer for helpful discussions. This research was conducted by the Australian Research Council Centre of Excellence for Quantum Computation and Communication Technology (project no. CE110001027) and the US National Security Agency and US Army Research Office (contract no. W911NF-08-1-0527). M.Y.S. acknowledges an ARC Laureate Fellowship.

\appendix
\section{Hamiltonian definition}
The Hamiltonian for exchange coupled donors including hyperfine, $H_{hf}$ and electrical detuning, $H_{\Delta}$ written in the singlet-triplet basis of the donor electrons is given by,
\begin{widetext}
\begin{equation}
\begin{split}
H_{ST} &=H_{ze} + H_{zn} + H_{t_c} + H_{\Delta} + H_{hf} \\
				H_{ze}&=\gamma_e B_0\left(\ketbra{T_{11}^+}{T_{11}^+} - \ketbra{T_{11}^-}{T_{11}^-}\right) + \frac{\gamma_e}{2} B_0\left(\ketbra{\uparrow_{01}}{\uparrow_{01}} - \ketbra{\downarrow_{01}}{\downarrow_{01}}\right)\\
				H_{zn}&=\gamma_n B_0(\ketbra{\Downarrow\Downarrow}{\Downarrow\Downarrow} - \ketbra{\Uparrow\Uparrow}{\Uparrow\Uparrow}) \\
				H_{t_c} &= \frac{t_c}{2}\left(\ketbra{S_{11}}{S_{02}} + \ketbra{S_{02}}{S_{11}}\right) \\
				H_{\Delta} &=  - \Delta \ketbra{S_{02}}{S_{02}}\\
				H_{hf} &= A_L\vec{S}_{L}^{(1,1)}\cdot\sum_{i_L}\vec{I}_{i_L} + A_R\vec{S}_{R}^{(1,1)}\cdot\sum_{i_R}\vec{I}_{i_R} + A_R \vec{S}_1\cdot\sum_{i_R}\vec{I}_{i_R},
\label{eq:Singlet-Triplet}
\end{split}
\end{equation}
\end{widetext}
where in the Zeeman term, $\vec{B_0}{=}(0,0,B_0)$ and the gyromagnetic ratios of the electron and nuclear spin states are $\gamma_e{=}28.024$ GHz/T and $\gamma_n{=}17.235$ MHz/T which will define the respective resonance conditions during either electron or nuclear spin excitation. Throughout the article we assume that the donors are in the \textit{weak coupling} regime where valley-orbit effects are negligible. 

The terms $\vec{S}_L^{(1,1)}$ and $\vec{S}_R^{(1,1)}$ are the electron spin operators for the (1,1) charge state  and $\vec{I}_{i_L}$ and $\vec{I}_{i_R}$ are the nuclear spin operators for P$_L$ and P$_R$ (with donors numbers $i_L{=}i_R{=}1$ for single donors). To account for tunneling out of the $\ket{S_{02}}$ state to the drain we have included the single electron states ${\ket{\uparrow_{01}}}$ and ${\ket{\downarrow_{01}}}$ which represent the spin states of the electron in the (0,1) charge state~\cite{koppens2007}. For these single electron states, the system has three particles, two nuclei and one electron spin such that the hyperfine coupling is only between the electron spin, $\vec{S}_1$, and the second donor nuclei, $\vec{I}_{i_R}$. By considering the two donor nuclei quantum mechanically we can obtain the full system spin dynamics.

In total our Hamiltonian is spanned by $\{E\}{\otimes}\{N\}$ where $\{N\}$ represents the nuclear subspace of states ${\{\ket{\Uparrow \Uparrow}}, {\ket{\Uparrow \Downarrow}}, {\ket{\Downarrow \Uparrow}}, \ket{{\Downarrow \Downarrow}}\}$ and $\{E\}$ the electron subspace consisting of  ${\{\ket{T_{11}^{i}}}, {\ket{S_{11}}}, {\ket{S_{02}}}, {\ket{\uparrow_{01}}}, {\ket{\downarrow_{01}}\}}$, where $i=+,0,-$. 

To include incoherent terms we make use of a technique commonly used in the ESR/NMR community~\cite{jeener1982,ernst1987,mayne2007}, which involves transforming from Hilbert space ($\mathcal{H}_n$) to a higher dimensional space---Liouville space, $\mathcal{L}_{n^2}$. The incoherent processes can then be gathered into the dissipator superoperator, $\hat{\hat{\mathcal{D}}}$ in $\mathcal{L}_{n^2}$ that can model non-trace preserving decoherence processes without the need of finding the operators in $\mathcal{H}_n$.

\section{Rotating wave approximation}
The ESR term in the Hamiltonian (Eq.~\ref{eq:Singlet-Triplet}) is
\begin{equation}
H_{ESR} = \gamma_e B_{ac}^{esr} \cos{(\Omega t)} (S_{L(x)}^{(1,1)} + S_{R(x)}^{(1,1)} + S_{1(x)}),
\end{equation}
where we chose a realistic $B_{ac}^{esr}{=}1$~mT as the microwave magnetic field strength~\cite{koppens2006,pla2012}, $\Omega$ its frequency and $\{S_{L(x)}^{(1,1)}$, $S_{R(x)}^{(1,1)}$,$S_{1(x)}\}$ are the electron $x$-spin operators for the (1,1) and (0,1) charge state respectively. To remove the time-dependence of the ESR field we can make a transformation from the laboratory frame to the rotating frame of the Larmor frequency ($\omega_e$) of the electron spins. This is simply a rotation about $z$ by an angle $\theta{=}\omega_e t$. We can then remove the faster rotating terms by making the rotating wave approximation. This transforms the Hamiltonian according to
\begin{equation}
H_{rot} = R_z^{\theta} H R_z^{\theta \dag} - R_z^{\theta} \frac{d R_z^{\theta \dag}}{dt}.
\end{equation}
This amounts to a two term addition to the Hamiltonian
\begin{equation}
H_{rot} = H + H_1 - F,
\end{equation}
where $H$ is the non-rotating Hamiltonian, $H_1{=}\gamma_e B_1 (S_{L(x)}^{(1,1)}{+}S_{R(x)}^{(1,1)}{+}S_{1(x)})$ and $F{=}\Omega \sigma_z$. $H_1$ contains the ESR driving terms and $F$ is the correction term due to the approximation ($B_1{=}B_{ac}^{esr}/2$ and  $\sigma_z$ is the $z$ operator for the entire Hilbert space).

\section{Liouville Space} 
To incorporate decoherence and relaxation into the time evolution of the density operator, a master equation approach can be used. The problem is treated in a higher-dimensional space, known as Liouville space, $\mathcal{L}_{n^2}$. The master equation in Liouville space is given by~\cite{limes2013},
\begin{equation}
\frac{d\ket{\rho}}{dt} = \frac{i}{\hbar}\hat{\hat{\mathcal{L}}} \ket{\rho} + \hat{\hat{\mathcal{D}}} \ket{\rho} = \hat{\hat{G}} \ket{\rho}.
\label{eq:Gdef}
\end{equation}
Importantly, the density operator is now a vector of length $n^2$, which is operated upon by the Liouvillian superoperator, $\hat{\hat{\mathcal{L}}}{=}\mathbb{I}{\otimes}H{-}H{\otimes}\mathbb{I}$ and the dissipator superoperator $\hat{\hat{\mathcal{D}}}$ representing incoherent terms. When $\hat{\hat{G}}$ is time-independent Eq.~\ref{eq:Gdef} has the solution,
\begin{equation}
\ket{\rho(t)} = e^{(\hat{\hat{G}}t)} \ket{\rho(0)}.
\end{equation}

\section{The dissipator superoperator}
The dissipator superoperator term contains the tunnel rates between the source and P$_L$, the drain and P$_R$, $\Gamma_L$ and $\Gamma_R$ respectively. The elements of $\hat{\hat{\mathcal{D}}}$ that contain these rates are determined by the effect they have on the system~\cite{koppens2005,koppens2007}. 

Two electron transport cycles were studied in this paper, for the first case: (0,1){$\rightarrow$}(0,2){$\rightarrow$}(1,1){$\rightarrow$}(0,1), the electrons are tunneling from the drain through the dots to the source. The diagonal dissipator elements in this scneario are,
\begin{equation}
\begin{aligned}
\hat{\hat{\mathcal{D}}}[\ket{T_{11}^+}] =& - \Gamma_L \ket{\rho_{T_{11}^+ T_{11}^+}}\\ 
\hat{\hat{\mathcal{D}}}[\ket{T_{11}^0}] =& - \Gamma_L \ket{\rho_{T_{11}^0 T_{11}^0}}\\ 
\hat{\hat{\mathcal{D}}}[\ket{S_{11}}] =& - \Gamma_L \ket{\rho_{S_{11} S_{11}}}\\ 
\hat{\hat{\mathcal{D}}}[\ket{T_{11}^-}] =& - \Gamma_L \ket{\rho_{T_{11}^- T_{11}^-}}\\ 
\hat{\hat{\mathcal{D}}}[\ket{S_{02}}] =& \Gamma_R (\ket{\rho_{\uparrow_{01} \uparrow_{01}}} + \ket{\rho_{\downarrow_{01} \downarrow_{01}}})\\ 
\hat{\hat{\mathcal{D}}}[\ket{\uparrow_{01}}] =& \Gamma_L \ket{\rho_{T_{11}^+ T_{11}^+}} + \frac{\Gamma_L}{2} (\ket{\rho_{T_{11}^0 T_{11}^0}} + \ket{\rho_{S_{11} S_{11}}})\\ &- \Gamma_R \ket{\rho_{\uparrow_{01} \uparrow_{01}}}\\ 
\hat{\hat{\mathcal{D}}}[\ket{\downarrow_{01}}] =& \Gamma_L \ket{\rho_{T_{11}^- T_{11}^-}} + \frac{\Gamma_L}{2} (\ket{\rho_{T_{11}^0 T_{11}^0}} + \ket{\rho_{S_{11} S_{11}}})\\ &- \Gamma_R \ket{\rho_{\downarrow_{01} \downarrow_{01}}},
\end{aligned}
\end{equation}
where $\ket{\rho_{jk}}$ indicates the the density operator element that $\hat{\hat{\mathcal{D}}}[j]$ acts upon. The off-diagonal elements between states $j$ and $k$ are
\begin{equation}
\hat{\hat{\mathcal{D}}}[\ket{jk}] = - \frac{\Gamma_{j,k}}{2}\ket{\rho_{jk}},
\label{eq:offD}
\end{equation}
here $\Gamma_{j,k}{=}\Gamma_R$ ($\Gamma_L$) if $j,k{=}\ket{\uparrow_{01}}$ or $\ket{\downarrow_{01}}$ ($\ket{T_{11}^i}$, $\ket{S_{11}}$); otherwise, it is zero. The tunneling rates account for the loss of coherence between the states $j$ and $k$ during transport and must be included to ensure positivity of the density operator.

The current through donors is given by the probability of the system to be in the (1,1) charge configuration multiplied by the tunnel rate from P$_L$ to the drain
\begin{align}
I &= |e| \Gamma_L (\ket{\rho_{T_{11}^+T_{11}^+}} + \ket{\rho_{T_{11}^0T_{11}^0}} + \ket{\rho_{T_{11}^-T_{11}^-}} + \ket{\rho_{S_{11}S_{11}}}) \nonumber \\
  &= |e| \Gamma_L (P(T_{11}^+) + P(T_{11}^0) + P(T_{11}^-) + P(S_{11})) \nonumber \\
  &= |e| \Gamma_L P\textnormal{(1,1)}.
\end{align}

For the electron transport cycle: (0,1){$\rightarrow$}(1,1){$\rightarrow$}(0,2){$\rightarrow$}(0,1), the electrons are tunneling from the source through the donors to the drain. So here the dissipator elements are essentially reversed
\begin{equation}
\begin{aligned}
\hat{\hat{\mathcal{D}}}[\ket{T_{11}^+}] =& \frac{\Gamma_L}{2}\ket{\rho_{\uparrow_{01}\uparrow_{01}}}\\
\hat{\hat{\mathcal{D}}}[\ket{T_{11}^0}] =& \frac{\Gamma_L}{4}(\ket{\rho_{\uparrow_{01}\uparrow_{01}}} +\ket{\rho_{\downarrow_{01}\downarrow_{01}}})\\
\hat{\hat{\mathcal{D}}}[\ket{S_{11}}] =& \frac{\Gamma_L}{4}(\ket{\rho_{\uparrow_{01}\uparrow_{01}}} +\ket{\rho_{\downarrow_{01}\downarrow_{01}}})\\
\hat{\hat{\mathcal{D}}}[\ket{T_{11}^-}] =& \frac{\Gamma_L}{2}\ket{\rho_{\downarrow_{01}\downarrow_{01}}}\\
\hat{\hat{\mathcal{D}}}[\ket{S_{02}}] =& - \Gamma_R\ket{\rho_{S_{02}S_{02}}}\\
\hat{\hat{\mathcal{D}}}[\ket{\uparrow_{01}}] =& \frac{\Gamma_R}{2}\ket{\rho_{S_{02}S_{02}}} - \Gamma_L\ket{\rho_{\uparrow_{01}\uparrow_{01}}}\\
\hat{\hat{\mathcal{D}}}[\ket{\downarrow_{01}}] =& \frac{\Gamma_R}{2}\ket{\rho_{S_{02}S_{02}}} - \Gamma_L\ket{\rho_{\downarrow_{01}\downarrow_{01}}},
\end{aligned}
\end{equation}
where now, for the off-diagonal elements, $\Gamma_{j,k}{=}\Gamma_R$ ($\Gamma_L$) if $j,k{=}\ket{S_{02}}$ ($\ket{\uparrow_{01}}$ or $\ket{\downarrow_{01}}$); otherwise, it is zero.

To determine the current through the donors we calculate the probability for the system to be in $\ket{S_{02}}$ multiplied by the tunnel rate from P$_R$ to the drain~\cite{jouravlev2006}
\begin{equation}
I = |e| \Gamma_R \ket{\rho_{S_{02}S_{02}}} = |e| \Gamma_R P(\ket{S_{02}}).
\end{equation}

Throughout the paper we choose $\Gamma_L{=}\Gamma_R{=}100$ MHz, such that it is less than $A$ so that the hyperfine interaction has time to mix the electron states but still large enough to give an appreciable current through the donors.

\newpage
$ $\\
\newpage
\bibliography{MyBib}

\begin{thebibliography}{44}
\providecommand{\natexlab}[1]{#1}
\providecommand{\url}[1]{\texttt{#1}}
\expandafter\ifx\csname urlstyle\endcsname\relax
  \providecommand{\doi}[1]{doi: #1}\else
  \providecommand{\doi}{doi: \begingroup \urlstyle{rm}\Url}\fi

\bibitem[Johnson et~al.(2005)Johnson, Petta, Marcus, Hanson, and
  Gossard]{johnson2005}
A.~C. Johnson, J.~R. Petta, C.~M. Marcus, M.~P. Hanson, and A.~C. Gossard.
\newblock Singlet-triplet spin blockade and charge sensing in a few-electron
  double quantum dot.
\newblock \emph{Nature}, 435:\penalty0 925--928, 2005.

\bibitem[Koppens et~al.(2005)Koppens, Folk, Elzerman, Hanson, {Williems van
  Beveran}, Vink, Tranitz, Wegscheider, Kouwenhoven, and
  Vandersypen]{koppens2005}
F.~H.~L. Koppens, J.~A. Folk, J.~M. Elzerman, R.~Hanson, L.~H. {Williems van
  Beveran}, I.~T. Vink, H.~P. Tranitz, W.~Wegscheider, L.~P. Kouwenhoven, and
  L.~M.~K. Vandersypen.
\newblock Control and detection of singlet-triplet mixing in a random nuclear
  field.
\newblock \emph{Science}, 309:\penalty0 1346--1350, 2005.

\bibitem[Koppens et~al.(2006)Koppens, Buizert, Tielrooij, Vink, Nowack,
  Meunier, Kouwenhoven, and Vandersypen]{koppens2006}
F.~H.~L. Koppens, C.~Buizert, K.~J. Tielrooij, I.~T. Vink, K.~C. Nowack,
  T.~Meunier, L.~P. Kouwenhoven, and L.~M.~K. Vandersypen.
\newblock Driven coherent oscillations of a single electron spin in a quantum
  dot.
\newblock \emph{Nature}, 442:\penalty0 766--771, 2006.

\bibitem[Nowack et~al.(2011)Nowack, Shafiei, Laforest, Prawiroatmodjo,
  Schreiber, Recihl, Wegscheider, and Vandersypen]{nowack2011}
K.~C. Nowack, M.~Shafiei, M.~Laforest, G.~E. D.~K. Prawiroatmodjo, L.~R.
  Schreiber, C.~Recihl, W.~Wegscheider, and L.~M.~K. Vandersypen.
\newblock Single-shot correlations and two-qubit gate of solid-state spins.
\newblock \emph{Science}, 333:\penalty0 1269--1272, 2011.

\bibitem[Petta et~al.(2005)Petta, Johnson, Taylor, Laird, Yacoby, Lukin,
  Marcus, Hanson, and Gossard]{petta2005}
J.~R. Petta, A.~C. Johnson, J.~M. Taylor, E.~A. Laird, A.~Yacoby, M.~D. Lukin,
  C.~M. Marcus, M.~P. Hanson, and A.~C. Gossard.
\newblock Coherent manipulation of coupled electron spins in semiconductor
  quantum dots.
\newblock \emph{Science}, 309:\penalty0 2180--2184, 2005.

\bibitem[Schuetz et~al.(2014)Schuetz, Kessler, Vandersypen, Cirac, and
  Giedke]{schuetz2014}
M.~J.~A. Schuetz, E.~M. Kessler, L.~M.~K. Vandersypen, J.~I. Cirac, and
  G.~Giedke.
\newblock Nuclear spin dynamics in double quantum dots: {Multistability,
  dynamical polarisation, criticality, and entanglement}.
\newblock \emph{Phys. Rev. B}, 89:\penalty0 195310, 2014.

\bibitem[Prati et~al.(2012)Prati, Hori, Guagliardo, Ferrari, and
  Shinada]{prati2012}
E.~Prati, M.~Hori, F.~Guagliardo, G.~Ferrari, and T.~Shinada.
\newblock Anderson-mott transition in arrays of a few dopant atoms in a silicon
  transistor.
\newblock \emph{Nature Nanotech.}, 7:\penalty0 443--447, 2012.

\bibitem[Gonzalez-Zalba et~al.(2014)Gonzalez-Zalba, Saraiva, Calderon, Heiss,
  Koiller, and Fergurson]{gonzalez-zalba2014}
M.~F. Gonzalez-Zalba, A.~Saraiva, M.~J. Calderon, D.~Heiss, B.~Koiller, and
  A.~J. Fergurson.
\newblock An exchange-coupled donor molecule in silicon.
\newblock \emph{Nano Lett.}, 14:\penalty0 5672--5676, 2014.

\bibitem[Fuechsle et~al.(2012)Fuechsle, Miwa, Mahapatra, Ryu, Lee, Warschkow,
  Hollenberg, Klimeck, and Simmons]{fuechsle2012}
M.~Fuechsle, J.~A. Miwa, S.~Mahapatra, H.~Ryu, S.~Lee, O.~Warschkow, L.~C.~L.
  Hollenberg, G.~Klimeck, and M.~Y. Simmons.
\newblock A single-atom transistor.
\newblock \emph{Nature Nanotech.}, 7:\penalty0 242--246, 2012.

\bibitem[Weber et~al.(2012)Weber, Mahapatra, Watson, and Simmons]{weber2012a}
B.~Weber, S.~Mahapatra, T.~F. Watson, and M.~Y. Simmons.
\newblock Engineering independent electrostatic control of atomic-scale
  ($\sim$4 nm) silicon double quantum dots.
\newblock \emph{Nano Lett.}, 12:\penalty0 4001--4006, 2012.

\bibitem[Buch et~al.(2013)Buch, Mahapatra, Rahman, Morello, and
  Simmons]{buch2013}
H.~Buch, S.~Mahapatra, R.~Rahman, A.~Morello, and M.~Y. Simmons.
\newblock Spin readout and addressability of phosphorous-donor clusters in
  silicon.
\newblock \emph{Nature Commun.}, 4:\penalty0 2017, 2013.

\bibitem[Hollenberg et~al.(2006)Hollenberg, Greentree, Fowler, and
  Wellard]{hollenberg2006}
L.~C.~L. Hollenberg, A.~D. Greentree, A.~G. Fowler, and C.~J. Wellard.
\newblock Two-dimensional architectures for donor-based quantum computing.
\newblock \emph{Phys. Rev. B}, 74:\penalty0 045311, 2006.

\bibitem[Bose(2003)]{bose2003}
S.~Bose.
\newblock Quantum communication through an unmodulated spin chain.
\newblock \emph{Phys. Rev. Lett.}, 91:\penalty0 207901, 2003.

\bibitem[Bose(2007)]{bose2007}
S.~Bose.
\newblock Quantum communication through spin chain dynamics: an introductory
  overview.
\newblock \emph{Contemp. Phys.}, 48:\penalty0 13--30, 2007.

\bibitem[Kay(2010)]{kay2010}
A.~Kay.
\newblock Perfect, efficient, state transfer and its applications as a
  constructive tool.
\newblock \emph{Int. J. Quantum Inf.}, 8:\penalty0 641, 2010.

\bibitem[Rahman et~al.(2009)Rahman, Park, Cole, Greentree, Muller, Klimeck, and
  Hollenberg]{rahman2009}
R.~Rahman, S.~H. Park, J.~H. Cole, A.~D. Greentree, R.~P. Muller, G.~Klimeck,
  and L.~C.~L. Hollenberg.
\newblock Atomistic simulations of adiabatic coherent electron transport in
  triple donor systems.
\newblock \emph{Phys. Rev. B}, 80:\penalty0 035302, 2009.

\bibitem[Rahman et~al.(2010)Rahman, Muller, Levy, Carroll, Klimeck, Greentree,
  and Hollenberg]{rahman2010}
R.~Rahman, R.~P. Muller, J.~E. Levy, M.~S. Carroll, G.~Klimeck, A.~D.
  Greentree, and L.~C.~L. Hollenberg.
\newblock Coherent electron transport by adiabatic passage in an imperfect
  donor chain.
\newblock \emph{Phys. Rev. B}, 82:\penalty0 155315, 2010.

\bibitem[Cirac and Zoller(2000)]{cirac2000}
J.~I. Cirac and P.~Zoller.
\newblock A scalable quantum computer with ions in an array of microtraps.
\newblock \emph{Nature}, 404:\penalty0 579--581, 2000.

\bibitem[Skinner et~al.(2003)Skinner, Davenport, and Kane]{skinner2003}
A.~J. Skinner, M.~E. Davenport, and B.~E. Kane.
\newblock Hydrogenic spin quantum computing in silicon: {A} digital approach.
\newblock \emph{Phys. Rev. Lett.}, 90:\penalty0 087901, 2003.

\bibitem[Loss and DiVincenzo(1998)]{loss1998}
D.~Loss and D.~P. DiVincenzo.
\newblock Quantum computation with quantum dots.
\newblock \emph{Phys. Rev. A}, 57:\penalty0 120--126, 1998.

\bibitem[{van der Wiel} et~al.(2003){van der Wiel}, {De Franceschi}, Elzerman,
  Fujisawa, Tarucha, and Kouwenhoven]{vanderwiel2003}
W.~G. {van der Wiel}, S.~{De Franceschi}, J.~M. Elzerman, T.~Fujisawa,
  S.~Tarucha, and L.~P. Kouwenhoven.
\newblock Electron transport through double quantum dots.
\newblock \emph{Rev. Mod. Phys.}, 75:\penalty0 1--22, 2003.

\bibitem[Taylor et~al.(2007)Taylor, Petta, Johnson, Yacoby, Marcus, and
  Lukin]{taylor2007}
J.~M. Taylor, J.~R. Petta, A.~C. Johnson, A.~Yacoby, C.~M. Marcus, and M.~D.
  Lukin.
\newblock Relaxation, dephasing, and quantum control of electron spins in
  double quantum dots.
\newblock \emph{Phys. Rev. B}, 76:\penalty0 035315, 2007.

\bibitem[Hanson et~al.(2007)Hanson, Kouwenhoven, Petta, Tarucha, and
  Vandersypen]{hanson2007}
R.~Hanson, L.~P. Kouwenhoven, J.~R. Petta, S.~Tarucha, and L.~M.~K.
  Vandersypen.
\newblock Spins in few-electron quantum dots.
\newblock \emph{Rev. Mod. Phys.}, 79:\penalty0 1217--1265, 2007.

\bibitem[Schliemann et~al.(2003)Schliemann, Khaetskii, and
  Loss]{schliemann2003}
J.~Schliemann, A.~Khaetskii, and D.~Loss.
\newblock Electron spin dynamics in quantum dots and related nanostructures due
  to hyperfine interaction with nuclei.
\newblock \emph{J. Phys.: Condens. Matter}, 15:\penalty0 R1809--R1833, 2003.

\bibitem[Muhonen et~al.(2014)Muhonen, Dehollain, Laucht, Hudson, Kalra,
  Sekiguchi, Itoh, Jamieson, McCallum, Dzurak, and Morello]{muhonen2014}
J.~T. Muhonen, J.~P. Dehollain, A.~Laucht, F.~E. Hudson, R.~Kalra,
  T.~Sekiguchi, K.~M. Itoh, D.~N. Jamieson, J.~C. McCallum, A.~S. Dzurak, and
  A.~Morello.
\newblock Storing quantum information for 30 seconds in a nanoelectronic
  device.
\newblock \emph{Nature Nanotech.}, 9:\penalty0 986--991, 2014.

\bibitem[Koiller et~al.(2002)Koiller, Hu, and {Das Sarma}]{koiller2002}
B.~Koiller, X.~Hu, and S.~{Das Sarma}.
\newblock Exchange in silicon-based quantum computer architecture.
\newblock \emph{Phys. Rev. Lett.}, 88:\penalty0 027903, 2002.

\bibitem[Wellard et~al.(2003)Wellard, Hollenberg, Pariosoli, Kettle, Goan,
  McIntosh, and Jamieson]{wellard2003}
C.~J. Wellard, L.~C.~L. Hollenberg, F.~Pariosoli, L.~M. Kettle, H.-S. Goan,
  J.~A.~L. McIntosh, and D.~N. Jamieson.
\newblock Electron exchange coupling for single-donor solid-state spin qubits.
\newblock \emph{Phys. Rev. B}, 68:\penalty0 195209, 2003.

\bibitem[Ernst et~al.(1987)Ernst, Bodenhausen, and Wokaun]{ernst1987}
R.~R. Ernst, G.~Bodenhausen, and A.~Wokaun.
\newblock \emph{Principles of Nuclear Magnetic Resonance in One and Two
  Dimensions}.
\newblock Oxford University Press, 1987.

\bibitem[Nazarov(1993)]{nazarov1993}
Yu.~V. Nazarov.
\newblock Quantum interference, tunnel junctions and resonant tunneling
  interferometer.
\newblock \emph{Physica B}, 189:\penalty0 57--69, 1993.

\bibitem[Jouravlev and Nazarov(2006)]{jouravlev2006}
O.~N. Jouravlev and Y.~V. Nazarov.
\newblock Electron transport in a double quantum dot governed by a nuclear
  magnetic field.
\newblock \emph{Phys. Rev. Lett.}, 96:\penalty0 176804, 2006.

\bibitem[Ono et~al.(2002)Ono, Austing, Tokura, and Tarucha]{ono2002}
K.~Ono, D.~G. Austing, Y.~Tokura, and Y.~Tarucha.
\newblock Current rectification by {Pauli} exclusion in a weakly coupled double
  quantum dot system.
\newblock \emph{Science}, 297:\penalty0 1313--1317, 2002.

\bibitem[Koppens et~al.(2007)Koppens, Buizert, Vink, Nowack, Meunier,
  Kouwenhoven, and Vandersypen]{koppens2007}
F.~H.~L. Koppens, C.~Buizert, I.~T. Vink, K.~C. Nowack, T.~Meunier, L.~P.
  Kouwenhoven, and L.~M.~K. Vandersypen.
\newblock Detection of single electron spin resonance in a double quantum dot.
\newblock \emph{J. Appl. Phys.}, 101:\penalty0 081706, 2007.

\bibitem[McCamey et~al.(2009)McCamey, {van Tol}, Morely, and
  Boehme]{mccamey2009}
D.~R. McCamey, J.~{van Tol}, G.~W. Morely, and C.~Boehme.
\newblock Fast nuclear spin hyperpolarization of phosphorus in silicon.
\newblock \emph{Phys. Rev. Lett.}, 102:\penalty0 027601, 2009.

\bibitem[McCamey et~al.(2010)McCamey, {van Tol}, Morely, and
  Boehme]{mccamey2010}
D.~R. McCamey, J.~{van Tol}, G.~W. Morely, and C.~Boehme.
\newblock Electronic spin storage in an electrically readable nuclear spin
  memory with a lifetime $>100$ seconds.
\newblock \emph{Science}, 330:\penalty0 1652--1656, 2010.

\bibitem[Weber et~al.(2014)Weber, Tan, Mahapatra, Watson, Ryu, Rahman,
  Hollenberg, Kilmeck, and Simmons]{weber2014}
B.~Weber, Y.~H.~Matthias Tan, S.~Mahapatra, T.~F. Watson, H.~Ryu, R.~Rahman,
  L.~C.~L. Hollenberg, G.~Kilmeck, and M.~Y. Simmons.
\newblock Spin blockade and exchange in {Coulomb}-confined silicon double
  quantum dots.
\newblock \emph{Nature Nanotech.}, 9:\penalty0 430--435, 2014.

\bibitem[Watson et~al.(2014)Watson, Weber, Miwa, Mahapatra, Heijen, and
  Simmons]{watson2014}
T.~Watson, B.~Weber, J.~A. Miwa, S.~Mahapatra, R.~M.~P. Heijen, and M.~Y.
  Simmons.
\newblock Transport in asymmetrically coupled donor-based silicon triple
  quantum dots.
\newblock \emph{Nano Lett.}, 14:\penalty0 1830--1835, 2014.

\bibitem[Bradbury et~al.(2006)Bradbury, Tyryshkin, Sabouret, Bokor, Schenkel,
  and Lyon]{bradbury2006}
F.~R. Bradbury, A.~M. Tyryshkin, G.~Sabouret, J.~Bokor, T.~Schenkel, and S.~A.
  Lyon.
\newblock Stark tuning of donor electron spins in silicon.
\newblock \emph{Phys. Rev. Lett.}, 97:\penalty0 176404, 2006.

\bibitem[Pla et~al.(2012)Pla, Tan, Dehollain, Lim, Morton, Jamieson, Dzurak,
  and Morello]{pla2012}
J.~J. Pla, K.~Y. Tan, J.~P. Dehollain, W.~H. Lim, J.~J.~L. Morton, D.~N.
  Jamieson, A.~S. Dzurak, and A.~Morello.
\newblock A single-atom electron spin qubit in silicon.
\newblock \emph{Nature}, 489:\penalty0 541--545, 2012.

\bibitem[Morello et~al.(2010)Morello, Pla, Zwanenburg, Chan, Tan, Huebl,
  Mottonen, Nugroho, Yang, {van Donkelaar}, Alves, Jamieson, Escott,
  Hollenberg, Clark, and Dzurak]{morello2010}
A.~Morello, J.~J. Pla, F.~A. Zwanenburg, K.~W. Chan, K.~Y Tan, H.~Huebl,
  M.~Mottonen, C.~D. Nugroho, C.~Yang, J.~A. {van Donkelaar}, A.~D.~C. Alves,
  D.~N. Jamieson, C.~C. Escott, L.~C.~L. Hollenberg, R.~G. Clark, and A.~S.
  Dzurak.
\newblock Single-shot readout of an electron spin in silicon.
\newblock \emph{Nature}, 467:\penalty0 687--691, 2010.

\bibitem[Kalra et~al.(2014)Kalra, Laucht, Hill, and Morello]{kalra2014}
R.~Kalra, A.~Laucht, C.~D. Hill, and A.~Morello.
\newblock Robust two-qubit gates for donors in silicon controlled by hyperfine
  interactions.
\newblock \emph{Phys. Rev. X}, 4:\penalty0 021044, 2014.

\bibitem[Oh et~al.(2013)Oh, Shim, Fei, Friessen, and Hu]{oh2013}
S.~Oh, Y.-P. Shim, J.~Fei, M.~Friessen, and X.~Hu.
\newblock Resonant adiabatic passage with three qubits.
\newblock \emph{Phys. Rev. A}, 87:\penalty0 022332, 2013.

\bibitem[Jeener(1982)]{jeener1982}
J.~Jeener.
\newblock Superoperators in magnetic resonance.
\newblock \emph{Adv. Magn. Opt. Res.}, 10:\penalty0 1--51, 1982.

\bibitem[Mayne(2007)]{mayne2007}
C.~L. Mayne.
\newblock Liouville equation of motion.
\newblock \emph{eMagRes.}, 2007.

\bibitem[Limes et~al.(2013)Limes, Wang, Baker, Lee, Saam, and
  Boehme]{limes2013}
M.~E. Limes, J.~Wang, W.~J. Baker, S.-Y. Lee, B.~Saam, and C.~Boehme.
\newblock Numerical study of spin-dependent transition rates within pairs of
  dipolar and exchange coupled spins with $s = \frac{1}{2}$ during magnetic
  resonant excitation.
\newblock \emph{Phys. Rev. B}, 87:\penalty0 165204, 2013.

\end{thebibliography}
\bibliographystyle{unsrtnat}

\end{document}